\documentclass[10pt,twocolumn,letterpaper]{article}

\usepackage{wacv}

\usepackage{graphicx}
\usepackage{amsmath}
\usepackage{amssymb}
\usepackage{mathtools}
\usepackage{booktabs}
\usepackage{multirow}

\usepackage[pagebackref,breaklinks,colorlinks]{hyperref}

\usepackage[capitalize]{cleveref}
\crefname{section}{Sec.}{Secs.}
\Crefname{section}{Section}{Sections}
\Crefname{table}{Table}{Tables}
\crefname{table}{Tab.}{Tabs.}

\usepackage[nolist]{acronym}
\begin{acronym}
\acrodef{ST}[ST]{Set Transformer}
\acrodef{b-ALL}[b-ALL]{b-Cell Lymphoblastic Leukemia}
\acrodef{MAE}[MAE]{masked autoencoder}
\acrodef{DL}[DL]{Deep Learning}
\acrodef{MRD}[MRD]{Measurable Residual Disease}
\acrodef{GNN}[GNN]{Graph Neural Networks}
\acrodef{GAT}[GAT]{Graph Attention Network}
\acrodef{FCM}[FCM]{Flow Cytometry}
\acrodef{KNN}[KNN]{k-Nearest Neighbor}
\acrodef{SOTA}[SOTA]{State of The Art}
\acrodef{AML}[AML]{Acute Myeloid Leukemia}
\end{acronym}

\def\wacvPaperID{2045} 
\def\confName{WACV}
\def\confYear{2024}

\begin{document}

\title{FATE: Feature-Agnostic Transformer-based Encoder for learning generalized embedding spaces in flow cytometry data}

\author{Lisa Weijler\\
TU Wien\\
{\tt\small lweijler@cvl.tuwien.ac.at}
\and
Florian Kowarsch\\
TU Wien\\
{\tt\small  florian.kowarsch@gmail.com}
\and
Michael Reiter\\
TU Wien\\
{\tt\small rei@cvl.tuwien.ac.at}
\and
Pedro Hermosilla\\
TU Wien\\
{\tt\small phermosilla@cvl.tuwien.ac.at}
\and
Margarita Maurer-Granofszky\\
St. Anna CCRI\\
{\tt\small margarita.maurer@ccri.at}
\and
Michael Dworzak\\
St. Anna CCRI\\
{\tt\small michael.dworzak@stanna.at}
}
\maketitle

\begin{abstract}

While model architectures and training strategies have become more generic and flexible with respect to different data modalities over the past years, a persistent limitation lies in the assumption of fixed quantities and arrangements of input features. This limitation becomes particularly relevant in scenarios where the attributes captured during data acquisition vary across different samples. 
In this work, we aim at effectively leveraging data with varying features, without the need to constrain the input space to the intersection of potential feature sets or to expand it to their union. We propose a novel architecture that can directly process data without the necessity of aligned feature modalities by learning a general embedding space that captures the relationship between features across data samples with varying sets of features. This is achieved via a set-transformer architecture augmented by feature-encoder layers, thereby enabling the learning of a shared latent feature space from data originating from heterogeneous feature spaces.
The advantages of the model are demonstrated for automatic cancer cell detection in acute myeloid leukemia in flow cytometry data,  where the features measured during acquisition often vary between samples. Our proposed architecture's capacity to operate seamlessly across incongruent feature spaces is particularly relevant in this context, where data scarcity arises from the low prevalence of the disease. The code is available for research purposes at \url{https://github.com/lisaweijler/FATE}.

\end{abstract}

\section{Introduction}
\label{sec:intro}

In recent years, a prominent trend in the machine learning landscape has been to design more generic architectures and learning strategies. Examples are multi-modality learning, data modality agnostic learning, or task agnostic learning. Successes in this field can be attributed to the rise of transformer models \cite{vaswani2017attention}, where (cross-)attention makes the combination of feature-embeddings of different modalities simple, as well as general pre-training and feature extraction methods such as the \ac{MAE}, where training does not depend on data modality or task-specific data augmentations. Yet, those approaches rely on unimodal encoders to process the different types of features corresponding to each data modality.

In this work, we propose a novel architecture, which is agnostic to the number and order of features of the input data by learning a generalized feature space using only one feature-agnostic encoder. We build upon the flexibility of transformers with respect to varying input sequence lengths and ordering. By using a feature encoding similar to the idea of positional encoding and an \ac{MAE} training strategy, the proposed model learns a general embedding space that captures the relationship between features across data samples. The common way of training an MAE is to mask parts of the input sample, such as patches of an image\cite{he2022masked}. In contrast, we propose masking of single feature values to learn the relationship between different features across datasets.

An application, where this characteristic holds significant relevance is the automated processing of \ac{FCM} data. 
An \ac{FCM} sample is essentially a set of feature vectors (events), each corresponding to a single cell. 
Based on the properties measured, diverse cell populations can be detected and analyzed. The choice of the features measured and used for analysis is highly task-dependent and often varies even within one use case due to different medical protocols, labs, and routines.
While this is no problem for manual analysis, where FCM samples are analyzed one by one, automated processing with e.g. deep learning requires a consistent set of features across datasets. 

We evaluate the proposed architecture on the task of cancer cell detection in pediatric \ac{AML} patients. The proportion of remaining cancer cells during and after treatment, called the \ac{MRD}, is an important factor for risk stratification and the development of effective treatment plans in accordance with individual patients’ needs. This is a particularly challenging task of \ac{FCM} data analysis, given the possible low proportions (down to $0.01\%$) and high heterogeneity of cancer cells due to patient-specific phenotype characteristics; sample- or patient-specific features are necessary to successfully distinguish healthy from cancerous cells. In addition, training data is scarce because of the low incidence of the disease (5-10 per mio. in Europe\cite{amlincidence}). In summary, our contribution is threefold:
\begin{enumerate}
    \item We introduce a novel architecture FATE (Feature-Agnostic Transformer-based Encoder), which, to the best of our knowledge, is the first one to enable processing of data with flexible input features. 
    \item We propose a pre-training strategy based on \ac{MAE} with a novel masking strategy that improves the quality of the representations learned by our model and it is agnostic to the number and type of features of the datasets.
    \item Using our proposed pre-training strategy we set a new \ac{SOTA} for the task of MRD detection in pediatric \ac{AML} patients and gain a performance increase of approximately $60\%$ compared to training from scratch. 
\end{enumerate}

\section{Related Work}
In this section, we describe the related work to this paper in different fields: Multimodal deep learning, \ac{MAE}, and \ac{FCM} data processing.

\paragraph{Multimodal deep learning} focuses on processing multiple input modalities such as images and text, to learn richer semantic features than unimodal data can offer, as well as for cross-modality tasks such as text-to-image generation. A prominent example is CLIP \cite{radford2021learning}, which is trained on text-image pairs to learn a cross-modality embedding used for different downstream tasks \cite{xu2023dream3d,wang2022clip}. Another pioneering work in this area is DALL-E \cite{ramesh2021zero} for text-to-image generation and several follow-up models. This is also a common practice in the medical field, where it enables the extraction of potentially complimenting features of diverse healthcare data modalities \cite{stahlschmidt2022multimodal, lin2022clustering, qiu2022multimodal, thakoor2022multimodal}. More recently ImageBind \cite{hou2023graphmae2} was proposed, which aims at learning a general embedding space for several data modalities such as audio, video, images and text. Those approaches, however, use several unimodal encoders or feature concatenation to process the features related to each modality and therefore expect a fixed number and ordering of features. In contrast, our model allows flexibility with respect to the features used. 

\paragraph{Masked autoencoding} has regained attention due to its capacity to learn useful embeddings from unlabelled data serving as an effective pre-training strategy for \ac{DL} models with an increasing number of parameters and capacities. \ac{MAE} are essentially a general form of denoising autoencoders \cite{vincent2008extracting} with a simple idea: mask i.e. remove a part of the input and let the model predict the missing part. This self-supervised training strategy does not depend on data-specific augmentations and is agnostic to the data modality. Starting from its success for natural language models such as BERT\cite{devlin2018bert}, it has made its way into vision \cite{he2022masked,feichtenhofer2022masked,woo2023convnext}, 3D data processing\cite{zhang2022point, jiang2023self,krispel2022maeli}, and many other domains\cite{hou2023graphmae2, majmundar2022met,Giri2020,baade2022mae}. \ac{MAE} also has been successfully employed in multimodal learning \cite{baade2022mae, gong2022contrastive,baevski2022data2vec}, yet, again they rely on different encoders for each data modality.

\paragraph{Automated FCM data processing} has a wide range of applications in a number of fields and is usually concerned with the detection and classification of novel or targeted cell populations. 
Early works of targeted cell population detection pool events of different samples from the training set together and train a classifier using pairs of single events and corresponding labels \cite{abdelaal2019LDA,Ni2016SVM,licandro2018wgan}. Methods using single events as input are restricted to learning fixed decision regions, while the relational positioning of cell populations to each other has proven as important information for successfully detecting rare or aberrant cells such as in MRD detection \cite{reiter2016clustering}. Therefore, approaches that process a whole FCM sample at once have emerged e.g. using Gaussian Mixture Models \cite{reiter2019automated} or transforming samples to images and applying convolutional neural networks\cite{Arvaniti2017cellcnn}. A more natural solution, without loss of information due to prior transformations are methods based on the attention mechanism \cite{vaswani2017attention,wodlinger2021automated, kowarsch2022xai}. Attention-based models are a way for event-level classification that learns and incorporates the relevance of other cell populations in a sample for the specific task. The current \ac{SOTA} for automated MRD detection \cite{wodlinger2021automated} is based on the \ac{ST} \cite{lee2019set}, an efficient variation of the Transformer model specifically designed for sets. Those approaches assume, however, a fixed set of features during training and inference. FCM data features are highly variable even within one dataset; reducing them to the greatest common intersection for consolidated processing dismisses discriminative information for the successful identification of targeted cells. There exists a work that aims at combining the features of samples by using nearest neighbor imputation\cite{leite2019cytobackbone,abdelaal2019cytofmerge,pedersen2022cycombine} but a recent study shows that those methods have severe limitations due to the questionable accuracy of imputed values for downstream analysis\cite{mocking2023merging}.

\section{Methods}
\label{sec:methods}
\begin{figure*}
  \centering
   \includegraphics[width=1.0\linewidth]{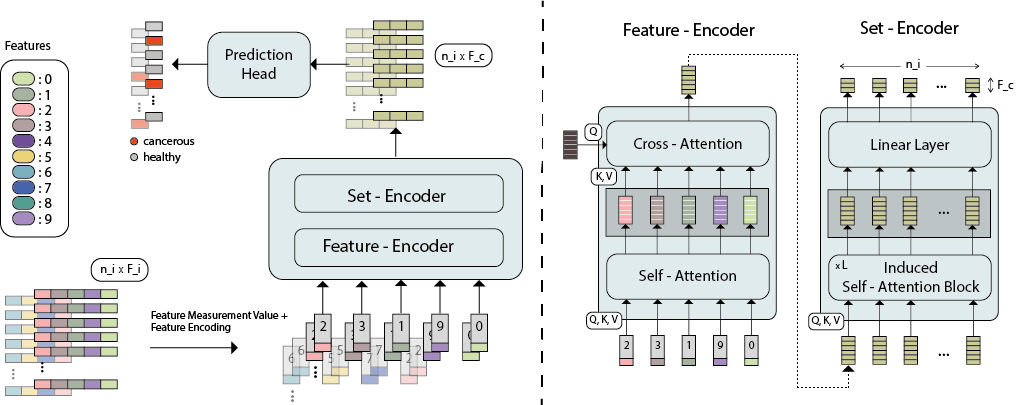}
   \caption{Illustration of the architecture of our model. \textbf{Left:} Each input feature of each event is concatenated with a positional encoding descriptor. These augmented features are updated using the other features of each event (Feature-Encoder) and additional events from the same sample (Set-Encoder).
   The resulting features, which are independent of the number and type of features used in the input, are used for the final prediction.
   \textbf{Right:} Detailed illustration of the Feature-Encoder and Set-Encoder networks.
   We use Self- and Cross-Attention between features in the Feature-Encoder and Induced Self-Attention between events within the sample.}
   \label{fig:fate-architecutre}
\end{figure*}

Our feature-agnostic model has a simple architecture depicted as an encoder and a prediction head. The transformer architecture is equivariant to permutations of the input sequence and can handle different sequence lengths. We exploit those properties in two ways. First, to be able to process events with different numbers and ordering of features, and second to be able to process samples with varying amounts of events. \cref{fig:fate-architecutre} shows the architecture described in \cref{subsec:FATencoder}.

For pre-training of our FATEncoder we use an \ac{MAE} that is trained to reconstruct the measurement values of masked features. Note that this differs from standard masking where whole instances of the data sample such as patches or pixels of images are masked. An illustration of the proposed \ac{MAE} is given in \cref{fig:fate-MAE}, and in \cref{subsec:mae-methods} its components are explained in detail. 

\subsection{Preliminaries}

Formally, we denote an input sample (set) $\boldsymbol{X}_i$ to our model as 
\begin{equation}
    \boldsymbol{X}_i = \{\boldsymbol{x}_{k_i}\in \mathbb{R}^{F_i}\}_{k_i = 1}^{n_i},
\end{equation}

where $F_i$ is the dimension of the feature vectors i.e. the number of features, and $n_i$ is the number of elements in the input set. One dataset constituted of $N$ samples is a set of sets 
\begin{equation}
    \{\boldsymbol{X}_i \in \mathbb{R}^{n_i \times F_i}\}_{i=1}^{N}.
\end{equation} 
In the context of \ac{FCM} data $\boldsymbol{x}_{k_i}$ is one event and $F_i$ the features measured by \ac{FCM} during acquisition. The number of events as well as the feature space can vary between samples; $n_i$ is typically between $10^4$ and $10^6$ and $F_i$ between 10 and 20 (limited by the properties of \ac{FCM} acquisition).

We use standard attention as proposed in \cite{vaswani2017attention} taking queries $\boldsymbol{Q}$, keys $\boldsymbol{K}$ and values $\boldsymbol{V}$ as input. For self-attention $\boldsymbol{Q}$, $\boldsymbol{K}$ and $\boldsymbol{V}$ are linear projections of the same input set  $\boldsymbol{X}_{i}$, while for cross-attention $\boldsymbol{Q}$ is either a learnable vector or our feature encodings as explained below, $\boldsymbol{K}$ and $\boldsymbol{V}$ are again linear projections of the same input set $\boldsymbol{X}_{i}$.

\subsection{FATEncoder}
\label{subsec:FATencoder}

The proposed encoder creates a common embedding space of fixed dimension $F_c$, regardless of the number and order of features of the input sample. It can be subdivided into two parts, which we call the \emph{Feature-Encoder} and the \emph{Set-Encoder}. The first part is responsible for being able to process inputs with different ordering and number of features, while the second part provides the context of all other events within one sample. 

In order to achieve the feature agnostic property, the Feature-Encoder treats the feature values of single events as input sequences, i.e. a sequence of scalar values. Given a sample $\boldsymbol{X}_i$, each event $\boldsymbol{x}_{k_i} \in \mathbb{R}^{F_i}$ thus turns into an input sequence 
\begin{equation}
    [x_{k_{i_1}},\dots,x_{k_{i_{F_i}}}]^\text{T} \in \mathbb{R}^{1 \times F_i}.
\end{equation}

We employ feature encoding similar to the idea of positional encoding, so the model knows what feature the scalar value in the sequence belongs to, and hence is able to relate them to others and learn their semantic meaning in the context of the dataset. The encoding is concatenated to the scalar measurement value of its corresponding feature, making the final input to our FATEncoder a sequence of vectors 
\begin{equation}
\begin{split}
    [\phi_E(\boldsymbol{x}_{k_i})_1,\dots,\phi_E(\boldsymbol{x}_{k_i})_{F_i}]^\text{T} \in \mathbb{R}^{1+D_{E} \times F_i},\\
     \phi_E(\boldsymbol{x}_{k_i})_j \coloneqq [x_{k_{i_j}}; \boldsymbol{E}_j] \in \mathbb{R}^{1+D_{E}},
\end{split}
\end{equation}

with $\boldsymbol{E}_j$ as the feature encoding vector for feature $j$ and $D_{E}$ as its dimension. 

Initially, self-attention is applied between those input vectors followed by cross-attention with a learned query vector $\boldsymbol{q} \in \mathbb{R}^{D_{\text{hidden}}}$ that summarizes the information of all features and transforms it into one single embedding vector of fixed dimension $\boldsymbol{z}_{k_i} \in \mathbb{R}^{D_{\text{hidden}}}$, 

\begin{equation}
    \begin{split}
     \boldsymbol{z}_{k_i} = \text{CrossAttn}(\boldsymbol{q}, \text{SelfAttn}(\phi_E(\boldsymbol{x}_{k_i})), \\
     \phi_E(\boldsymbol{x}_{k_i})\coloneqq
    [\phi_E(\boldsymbol{x}_{k_i})_1,\dots,\phi_E(\boldsymbol{x}_{k_i})_{F_i}].
     \end{split}
\end{equation}
After all elements of one sample have passed through the Feature-Encoder\footnote{A full sample $\boldsymbol{X}_i \in \mathbb{R}^{n_i \times F_i}$ can be efficiently processed as batch by using a tensor of dimension $n_i \times F_i \times 1+D_{E}$ as input.}, the Set-Encoder puts the event-embeddings $\boldsymbol{z}_{k_i}$ of one sample in context to each other. For this we utilize the Induced Self-Attention Blocks (ISABs) as introduced with the \ac{ST}\cite{lee2019set} (see supplementary materials). The ISAB uses learned queries (induced points) to reduce the $\mathcal{O}(n^2)$ complexity of self-attention to $\mathcal{O}(nm)$ with $n$ denoting the input sequence length and $m$ the number of induced points. Finally, a linear layer maps the embedding to its final embedding dimension $F_c$. The Set-Encoder can be summarized as,
\begin{equation}
    \{\boldsymbol{z}_{k_i}\}_i^{n_i} \leftarrow \text{LinLayer}(\text{ISAB}^{(l)}(\{\boldsymbol{z}_{k_i}\}_i^{n_i}),
\end{equation}
with $l$ being the number of ISAB layers applied.
\paragraph{Feature encoding}
In contrast to the positional encoding used in e.g. natural language processing, where ordinal encodings i.e. encodings that have a ranked ordering are used, our feature encoding should be nominal, thus not imposing an artificial relationship between values. A typical choice is one-hot encoding. However, with this choice, the dimension grows with the number of features covered and after training this number is locked-in given that adding new features would increase the dimension of the input vectors. This is why we chose learned encoding vectors of fixed dimension as used in \cite{dosovitskiy2020image} i.e. a learnable function
\begin{equation}
    e: \{0, \dots, M-1\} \mapsto \mathbb{R}^{D_E},
\end{equation}
where $M$ denotes the number of different features present in the training data set. 

See supplementary materials for a full list of features used.

\paragraph{Prediction head}
As prediction head, we use a 2-layer MLP with GELU as an activation function and a hidden layer dimension equal to the common feature space dimension $F_c$.

\subsection{FATE-\ac{MAE}}
\label{subsec:mae-methods}

\begin{figure}
  \centering
   \includegraphics[width=0.9\linewidth]{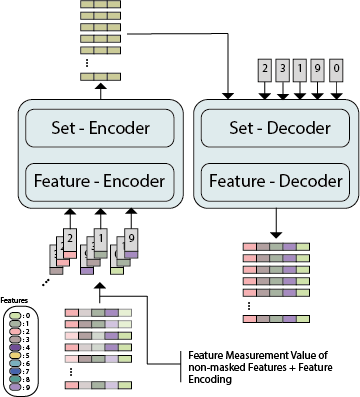}
   \caption{Masked autoencoder strategy used to train our model.
   During pre-training we mask a set of features for each event.
   The features remaining after the masking are given as input to our model to be encoded in a unified feature space.
   A decoder with a symmetric architecture of our encoder is used to predict the masked features in the initial step.}
   \label{fig:fate-MAE}
\end{figure}

We propose an \ac{MAE} strategy for pre-training our model. Like all autoencoders, our method uses an encoder that transforms the observed signal into a latent representation, and a decoder that reconstructs the original signal from this latent representation. However, our encoder is specifically designed to be able to process varying input features as described in \cref{subsec:FATencoder}. Consequently, our decoder has to be able to reconstruct a varying amount of features. The masking strategy and decoder architecture are described in the following; \cref{fig:fate-MAE} gives an overview of the \ac{MAE}. A detailed illustration of our proposed decoder architecture is given in the supplementary materials.

\paragraph{Masking}
Contrary to common masking strategies, where full instances e.g. patches of an image are masked, we mask a proportion of feature measurement values per element of the input set $\boldsymbol{X}_i$ i.e. per feature measurement vector. The proposed masking strategies allow us to learn the relationship between features in the context of the composition of cell populations within and across samples, in contrast to learning the relationship between cell populations given a fixed set of features, which is standard masking. The strategy is straightforward: We define a fixed ratio of features to mask (i.e. remove); a masking ratio of $0.5$ would mean removing half of the feature measurement values in each feature measurement vector of all samples. What features are to be masked is chosen randomly following a uniform distribution and can be different for each $\boldsymbol{x}_{k_i} \in \boldsymbol{X}_i$.

\paragraph{Decoder}
The decoder is symmetric to the encoder. Following the structure of the encoder, we divide it into the \emph{Set-Decoder} and \emph{Feature-Decoder}. The Set-Decoder consists of ISAB layers to capture the relation between the embeddings of one sample $\{\boldsymbol{z}_{k_i}\}_i^{n_i}$. The Feature-Decoder reconstructs the original feature measurement values of each set element $\boldsymbol{x}_{k_i} \in \mathbb{R}^{F_i}$ from its corresponding embedding vector  $\boldsymbol{z}_{k_i} \in \mathbb{R}^{F_c}$ given the context of all elements in the input set. This is not trivial due to the varying number of features between samples. To achieve this we condition the decoder on the features of which we want their measurement values to be reconstructed by cross-attending their feature encodings with the output of the set-decoder. Finally, similar to the feature-encoder we apply self-attention between the reconstructed feature measurement value vector of each $\boldsymbol{x}_{k_i}$ and map them with a linear layer to a scalar value. For a detailed illustration of the proposed decoder see supplementary materials.

\section{Experimental Setup}
In this section, we give an overview of the experiments conducted, the implementation details, and the datasets used. 

We focus on the task of \ac{MRD} detection of FCM samples in pediatric \ac{AML} patients given that it is one of the most challenging tasks in FCM data analysis with increased varying features between samples. \ac{AML} depicts a particularly high heterogeneity of cancer cells in comparison to other sub-types such as \ac{b-ALL}. Consequently, the features necessary to successfully separate healthy from cancer cells are highly diverse. In addition, it is a very rare disease resulting in a natural data scarcity. Hence, the ability to use other datasets for pre-training while being flexible with respect to input features is crucial.

We treat the problem as a binary classification, cancerous or healthy, and use binary cross-entropy as loss function for supervised training. Precision $p$, recall $r$, and $F$-score $F_1$ with cancer cells as positives are used as evaluation metrics. The metrics are calculated per sample, then averaged to determine the final metrics of the test set.

\subsection{Datasets}
The main data set (\emph{AML-MRD}) for the downstream task of MRD detection consists of 71 FCM samples coming from 12 different pediatric \ac{AML} patients. The samples were collected at different time points during therapy between the years 2021 and 2022. The number of markers used for samples of this dataset varies between 10 and 14. Additionally, the forward- and side-scatter of the laser light, which reflect the physical properties of the cells, are used as features. Between all samples of this dataset, 6 markers are shared. For a full listing of the number of samples per patient and the corresponding features see the supplementary materials.

For pre-training we use data from pediatric b-ALL patients at day 15 of therapy, data from pediatric AML patients at diagnosis, where cancerous cells have replaced most of the healthy cells, as well as FCM samples from screening cases or recovered patients, that are free of cancer cells (MRD negative). The b-ALL data is a publicly available\footnote{\href{https://flowrepository.org/id/FR-FCM-ZYVT}{flowrepository.org}} dataset (\emph{ALL-MRD}) collected from three different laboratories in Berlin, Buenos Aires, and Vienna. The diagnosis (\emph{DIA}) as well as the MRD negative dataset (\emph{CONTROL}) are a collection of samples from different laboratories in Vienna, Padua and Essen between the years 2016-2022. CONTROL and DIA have 4 features with AML-MRD in common, while ALL-MRD only has 2 excluding forward- and side-scatter. Supplementary materials give an overview of all features and the overlaps between datasets. 

Sampling and research were approved by local Ethics Committees, and informed consent was obtained from patients' or patients' parents or legal guardians according to the Declaration of Helsinki. Ground truth was obtained using manual analysis by at least two experts.
Whenever available, results were confirmed using an independent molecular methodology (RT-PCR).

The datasets we use for pre-training are not suitable as training data for the downstream task of MRD detection in pediatric AML patients, given the different types of cancer cells (ALL-MRD), composition of data (DIA), or lack of cancer cells (CONTROL). However, they can be successfully utilized for pre-training by our proposed model as we show in \cref{sec:results}. 

\cref{tab: data} provides a tabular overview of the data sets.

\begin{table}
    \centering
    \caption{Description of the FCM data sets used.}
    \label{tab: data}
    \begin{tabular}{cccc}
        \toprule
        {\bf Name} & {\bf City} & {\bf Years} & {\bf Samples} \\ [0.5ex] 
        \hline 
        {AML-MRD} & Vie  & 2021-22 & 71 \\
        \midrule
        {CONTROL} & Vie, Pad, Ess & 2016-21 & 308 \\ 
        {DIA} & Vie, Pad, Ess & 2016-22 & 110 \\ 
        {ALL-MRD} & Vie, Bln, Bue & 2009-14 & 338 \\ 

        \bottomrule
    \end{tabular}
\end{table}

\subsection{Experiments}

We conduct several experiments to evaluate the proposed FATE architecture and \ac{MAE}. The current \ac{SOTA} for automated MRD detection in FCM data based on the \ac{ST} \cite{wodlinger2021automated} serves as a baseline using the official implementation\footnote{\href{https://github.com/mwoedlinger/flowformer}{https://github.com/mwoedlinger/flowformer}}. 

We train the models from scratch and compare them to two different types of pre-training. First, we pre-train the baseline and the FATE model supervised with the \emph{CONTROL}, \emph{DIA}, and if applicable \emph{ALL-MRD} datasets. Second, we test our proposed pre-training method based on \ac{MAE} using the \emph{CONTROL}, \emph{DIA} and \emph{ALL-MRD} datasets. 

\paragraph{Implementation details}
For the training-from-scratch experiments, we train our model for 400 epochs and use early stopping after 300 epochs if there is no improvement of the $F_1$-score on the validation set. The baseline model is trained for 200 epochs with early stopping after 100 epochs since it showed a tendency to overfit on the small datasets. 

For the supervised pre-training we use 1500 epochs and finetune for 100 epochs using the pretained weights as initialization. Given that the task of finetuning is the same as for pre-training we do not remove the prediction head. For pre-training of the baseline we conduct two experiments. First, we only use the 4 features plus forward- and side-scatter that the CONTROL and DIA  have in common with the AML-MRD dataset and second, we use the union over all occuring features in all three datasets as input imputing the missing ones with zeros. For the FATE model each input sample's individual features can be utilized. 

Pre-training of the FATE \ac{MAE} is conducted for 2000 epochs for all experiments and pre-trained weights are again used as initialization for training of the downstream task. The model is finetuned for 300 epochs with early stopping after 200 and different masking-ratios are evaluated. 

The $L_1$-loss is used and calculated between the true measurement values of masked features and those reconstructed by the decoder. 
With respect to the learning rate for finetuning and from-scratch experiments, we opt for the cosine-annealing learning rate scheduler with a starting value of $0.001$, a minimum value of $0.0002$, and a maximum of 10 iterations. Throughout all pre-training experiments, we use the same scheduler with a starting value of $0.001$, a minimum value of $0.00002$, and a maximum of 100 iterations. AdamW optimizer is employed for all experiments.
A batch size of 32 and 8 is used for pre-training and finetuning, respectively. Training is conducted on an NVIDIA GeForce RTX 3090. 

All models are trained at least three times with different initialization and mean and standard deviations are reported.

\paragraph{Patient cross-validation} Since the target dataset \emph{AML-MRD} is small with 71 samples and several samples can originate from the same patient we employ a cross-validation to maximize the amount of training and evaluation data. The model should be capable of detecting MRD in FCM data of new patients, meaning for a fair assessment, samples taken from one patient must not be split up between the training, evaluation, and test set. Therefore we conducted a ``patient-cross-validation'', where each patient generates one split, i.e. the patient's samples are held out as test data. All remaining samples from other patients are divided into training and evaluation sets with a ratio of $0.8$ and $0.2$, respectively. The reported metrics are the average of the resulting values over all samples. See supplementary materials for a full list of splits and the number of training, validation, and test samples.

\paragraph{Model architecture}
For the baseline model we use the same architecture as proposed in \cite{wodlinger2021automated}, namely 4 ISAB layers with a hidden dimension of 32 and 16 induced points. For our FATE model, we use the architecture as described in \cref{sec:methods} with 3 ISAB layers in the encoder and decoder, a hidden dimension of $32$, and an embedding space dimension $F_c=8$. The learned feature encodings have a dimension of $D_E = 10$.

\begin{table*}
\caption{Results of our main experiments. We compare our model trained from scratch and pre-trained with different strategies to the current state-of-the-art Set-Transformer model~\cite{wodlinger2021automated}. The results show that our model performs similarly to ST on a supervised setup, but, when pre-trained with additional data, is able to outperform this model by using more data during pre-training.}
\setlength{\tabcolsep}{4pt}
\begin{center}
\begin{tabular}{lccccc}
\toprule
 &Pre-train & Dataset& ${p}$ &  ${r}$ & ${F_1}$\\
\midrule
\multirow{2}{*}{{ST}~\cite{wodlinger2021automated}} &-&-& 0.385 & 0.366&  0.314 {\scriptsize $\pm 0.015$}  \\
 &Sup.&CON, DIA& 0.446 & 0.356&  0.324 {\scriptsize $\pm 0.029$}  \\
 \midrule
 \multirow{3}{*}{{ST}~\cite{wodlinger2021automated} ({\scriptsize All Features})} &-&-& 0.328 & 0.456&  0.328 {\scriptsize $\pm 0.011$}  \\	
  &Sup.&CON, DIA& 0.493	 & 0.436&  0.332 {\scriptsize $\pm 0.017$}  \\

  &Sup.&CON, DIA, ALL-MRD& 0.457 & 0.385&  0.297 {\scriptsize $\pm 0.001$}  \\
 \midrule
\multirow{4}{*}{{FATE ({\scriptsize Ours})}}  &-&-&0.445& 0.435&   0.315 {\scriptsize $\pm 0.004$}  \\
                            &Sup.&CON, DIA&0.6175& 0.401&   0.389 {\scriptsize $\pm 0.058$}  \\
                                &Sup.&CON, DIA, ALL-MRD&0.531& 0.444&  {\bf 0.413} {\scriptsize $\pm 0.053$} \\
                                 &Sup.+MAE&CON, DIA, ALL-MRD&0.611& 0.550&  {\bf 0.496} {\scriptsize $\pm 0.045$}\\

\bottomrule
\end{tabular}
\vspace{-0.4cm}
\label{tab:results}
\end{center}
\end{table*}

\begin{table}
\caption{Results for different masking ratios of the input features. We can see that a strong masking ratio decreases performance due to the small number of features of each event.}
\setlength{\tabcolsep}{4pt}
\begin{center}
\begin{tabular}{lccc}
\toprule
 Masking ratio & ${p}$ &  ${r}$ & ${F_1}$\\
\midrule
$0.75\%$ & 0.574	 & 0.389 &  0.363 {\scriptsize $\pm 0.004$}  \\	
$0.5\%$ & 0.578 & 0.472&  0.459 {\scriptsize $\pm 0.016$}  \\		
$0.25\%$ &0.611& 0.550&  {\bf 0.496} {\scriptsize $\pm 0.045$}\\

\bottomrule
\end{tabular}
\vspace{-0.4cm}
\label{tab:masking-ratio}
\end{center}
\end{table}

\section{Results}
\label{sec:results}

\cref{tab:results} shows the results of the experiments conducted. Training from scratch yields similar results for all three setups, \ac{ST} with the intersection of all features, \ac{ST} with the union over all features in all datasets, and our FATE model. The \ac{ST} using all features as input, where missing ones are imputed with $0$, performs slightly better showing the benefit of sample-specific features. Our FATE model needs to learn the feature encoding and the common embedding space, which is hard given the little training set when training from scratch. Yet it holds comparable results with higher mean precision and recall. 

Pre-training the models supervised with the CON and DIA dataset improves results for all models. The biggest increase can be seen for our FATE model, indicating that the proposed architecture can successfully learn the relation between features across different samples. Surprisingly, when adding the ALL-MRD dataset to pre-training, the performance of the \ac{ST} drops. One explanation is the increased difference in features and in cancer cells to be detected since \ac{b-ALL} is a different leukemia subtype than \ac{AML}. For our proposed model the performance increases further despite those differences to a mean $F_1$-score of $0.413$, which can be interpreted as that the general relationship between markers in the context of the downstream task is extracted and beneficial for the model.

Pre-training with the proposed \ac{MAE} strategy is only possible for the FATE architecture. We can see that the results further increase to  $F_1 = 0.496$, yielding an improvement to approximately $60\%$ over training from scratch. The results indicate that our FATE model can successfully learn a common embedding space and utilize the information of all sample-specific features. Visualizations of the learned embeddings are provided in the supplementary material.

We evaluated different masking ratios for pre-training: $0.25\%$, $0.5\%$, and $0.75\%$. \cref{tab:masking-ratio} shows the results for those experiments. We can see that performance increases with a decreasing ratio, where $0.25\%$ yields the best performance. While the performance for a masking ratio of $0.5\%$ is still significantly better than supervised pre-training, when masking with a ratio of $0.75\%$, although still better than training from scratch, performance drops below supervised pre-training due to the small number of features for each event.

\section{Conclusion}
In this work, we introduce a novel architecture, FATE, that is agnostic to the number and order of features of the input data by learning a generalized feature space using only one feature-agnostic encoder. Further, we propose a pre-training strategy based on \ac{MAE} which, contrary to standard practices where complete feature vectors are masked during pre-training, masks part of the input features and tries to reconstruct them from the remaining ones. 
We show that, with this model and pre-training strategy, we are able to leverage new datasets during pre-training and improve current state-of-the-art on \ac{FCM} data by a large margin.

In the future, we would like to investigate which features are more relevant for feature reconstruction and downstream tasks by analyzing the attention weights of our model, in order to improve future data acquisition.
Moreover, we would like to investigate how the quality of the representations is affected by the number of common features during pre-training.

{\small
\bibliographystyle{ieee_fullname}
\bibliography{egbib}
}

\appendix

\section{Datasets}

\cref{tabsup: data} gives a detailed overview of the number of patients and their samples collected per dataset, as well as city and years of acquisition. All samples from the datasets used for pre-training (CONTROL, DIA, ALL-MRD) are from different patients than those in the target dataset, AML-MRD, for a fair assessment of generalization between patients.

\begin{table*}[h]
    \centering
    \caption{Description of the FCM datasets with respect to the city of the laboratory and the time-span the samples where acquired, the number of samples per dataset as well as the number of patients the samples were taken from.}
    \label{tabsup: data}
    \begin{tabular}{ccccc}
        \toprule
        {\bf Dataset} & {\bf City} & {\bf Years} & {\bf Samples} & {\bf Patients}\\ [0.5ex] 
        \hline 
        AML-MRD & Vienna  & 2021-2022 & 71 & 12\\
        \midrule
        \multirow{3}{*}{CONTROL} & Essen & 2016 & 52 & 17\\ 
         & Padua & 2016-2021 & 88 & 22 \\ 
         & Vienna & 2016-2021 & 168 & 57\\
         \midrule
       \multirow{3}{*}{DIA} & Essen & 2016 & 24 &12\\ 
        & Padua & 2016 & 6 & 3\\ 
         & Vienna  & 2016-2022 & 80 &30\\
        \midrule
        \multirow{3}{*}{ALL-MRD} & Vienna & 2009-2014 & 200 & 200\\ 
         & Berlin & 2015 & 72 & 72\\ 
         & Buenos Aires & 2016-2017 & 66 &66 \\
        \bottomrule
    \end{tabular}
\end{table*}

\begin{figure*}[h]
  \centering
   \includegraphics[width=0.9\linewidth]{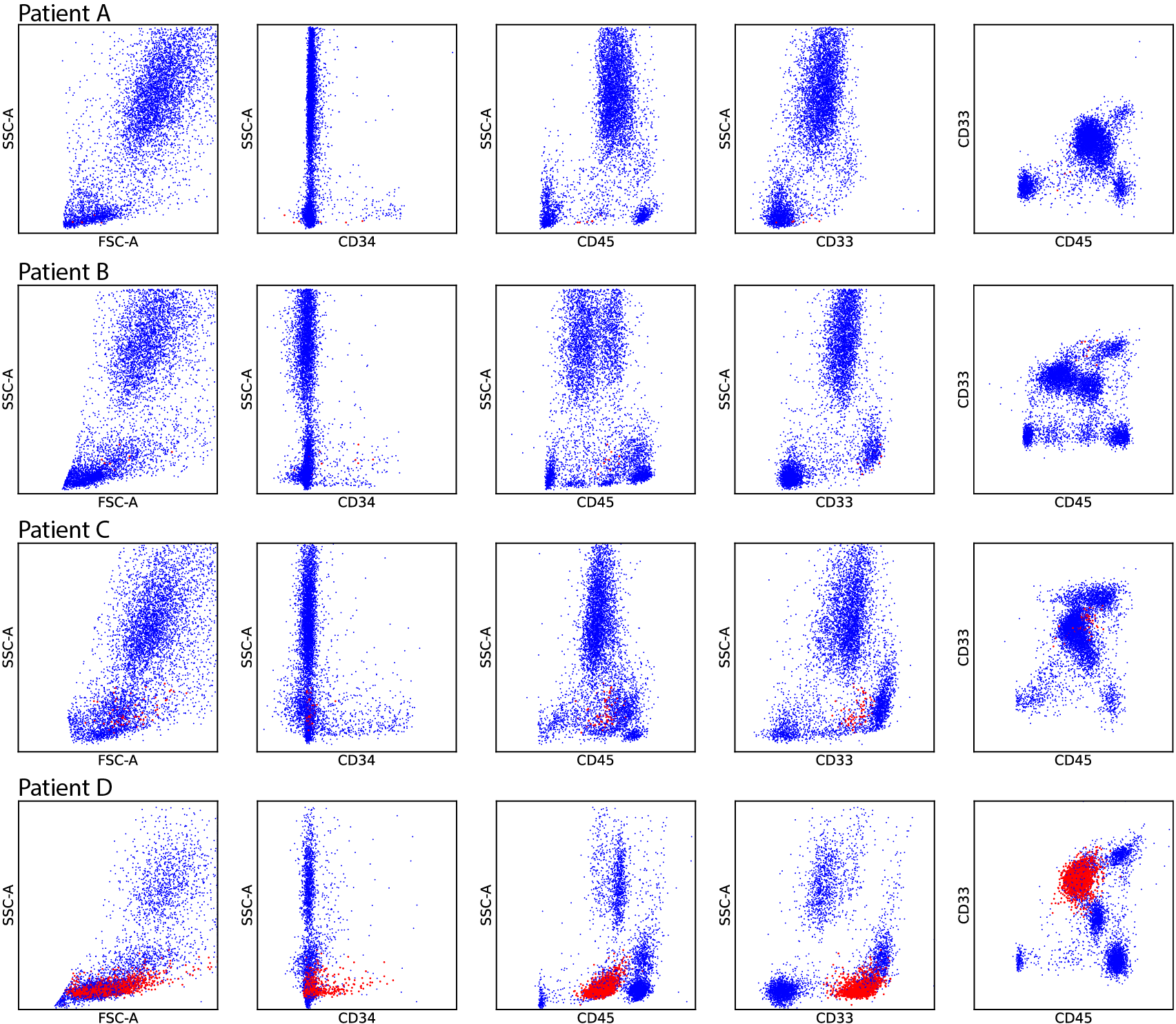}
   \caption{This figure shows 2D projections of samples from the AML-MRD dataset on different combination of features. Each dot corresponds to one event i.e. the feature measurement vector of one cell. Cancer cells are marked in red, healthy in blue. Each row corresponds to a different patient.}
   \label{fig:panels}
\end{figure*}

\cref{fig:panels} shows four different FCM samples of different patients coming from the AML-MRD dataset. The plots are 2D projections of the sample on different features including forward- and side-scatter (see \cref{supsec:Features}). Each point corresponds to the feature measurement vector of one cell. We can see the variation in cancer cells (red) regarding positioning and proportion as well as the variation in density of healthy cell populations (blue). 

\section{Features}
\label{supsec:Features}
All four datasets combined have 35 features in total including forward- (FSC) and side-scatter (SSC) of the laser light. FSC and SSC measure physical properties of the cell, like size and granularity. The rest of the features correspond to fluorescent-labeled surface markers, of which the concentration is measured. Different cell types bind to different surface markers and hence, those markers enable to separate and analyse cell-types. A detailed listing of what features in how many samples are present for each dataset is given in \cref{tab:features}. "CD" stands for cluster of differentiation. "A", "H" an "W" stands for area, hight and width, respectivley in the FSC and SSC. We just speak of FSC and SSC and do not necessarily differentiate between area, hight and width, since those are highly correlated. TIME indicates when the feature vector was acquired throughout the acquisition process of one sample and hence is not a strong discriminative feature. In total we can see that the AML-MRD and ALL-MRD datasets have the least in common  with only 2 discriminative features excluding FSC and SSC throughout all samples, CD45 and CD34.   
\begin{table*}
\centering
	\caption{List of features (markers) present in the datasets and their number of occurrences per dataset.}
	\label{tab:features}
	\begin{tabular}{lcccc}
		\toprule
		 {\bf Feature}  &  {\bf AML-MRD}  &  {\bf CONTROL} &  {\bf DIA} &  {\bf ALL-MRD} \\ [0.5ex]
		\hline
		FSC-A&71 &308& 110&338\\
        FSC-H& 66&131 & 58&0\\
        FSC-W& 71& 253& 99&338\\
        SSC-A& 71& 308& 110&338\\
        SSC-W& 67& 185&52 &1\\
        CD38& 32& 164& 52&338\\
        CD371& 32& 163& 52&0\\
        CD34& 71& 308& 110&338\\
        CD117& 71& 308& 110&0\\
        CD33& 71& 308& 110&0\\
        CD71& 5& 6& 28&0\\
        CD123& 33& 165& 53&0\\
        CD45RA& 32& 164& 52&0\\
        HLA-DR& 71&308 & 26&0\\
        CD45& 71& 308& 110&338\\
        TIME& 71& 308& 110&338\\
        CD99& 34& 180& 63&0\\
        CD10& 2& 0&12 &338\\
        CD133& 3& 0& 8&0\\
        CD15& 39& 161& 57&0\\
        CD11A& 18&11& 13&0\\
        CD7CD56& 5& 1& 3&0\\
        CD14& 39& 161& 57&0\\
        CD11B& 39& 161& 57&0\\
        CD13& 14& 132& 35&0\\
        NG2& 3& 2& 4&0\\
        CD56&12 &0 & 0&0\\
        CD7& 15& 15&11 &0\\
        CD312& 5& 0& 1&0\\
        CD16& 2&0 & 14&0\\
        CD48& 1& 0&2 &0\\
        CD58& 0& 0& 0&138\\
        CD19& 0& 3&3 &338\\
        CD20& 0& 0&0 &338\\
        SY41& 0& 0& 0&338\\
		\bottomrule
	\end{tabular}
\end{table*}

\section{Patient Cross-Validation}
The number of samples used for training, validation and testing are given in \cref{tab:datasplit}. Each patient determines one split. After precision $p$, recall $r$ and $F_1$-score have been determined for each sample in each test split, metrics are averaged over all samples to get the final results that are reported. 

\begin{table*}
\centering
	\caption{The sample of the AML-MRD dataset are divided into training-, evaluation-, and test-set. Each patient's samples exclusively form the test set, as indicated by the rightmost column specifying the sample count for each patient. The rest of the samples from different patients are allocated to the training and evaluation sets with a proportion of approximately $0.8$ and $0.2$, respectively.}
	\label{tab:datasplit}
	\begin{tabular}{lccc}
		\toprule
		 {\bf Patient}&  {\bf Train}   &  {\bf Eval}  &  {\bf Test} \\ [0.5ex]
		\hline
		A & 56 & 14 & 1 \\
	
		B & 51 & 15 & 5 \\

		C & 47 & 14 & 10 \\

		D & 54 & 13 & 4 \\
	
		E & 53 & 13 & 5 \\
	
		F & 49 & 11 & 11 \\

		G & 42 & 11 & 18 \\

		H & 56 & 13 & 2 \\

		I & 54 & 15 & 2 \\
	
		J & 51 & 11 & 9 \\
     
		K  & 56 & 13 & 2\\
   
		L & 54 & 15 & 2 \\
		\bottomrule
	\end{tabular}
\end{table*}

\section{FATE-Decoder}
\begin{figure*}[ht]
  \centering
   \includegraphics[width=0.45\linewidth]{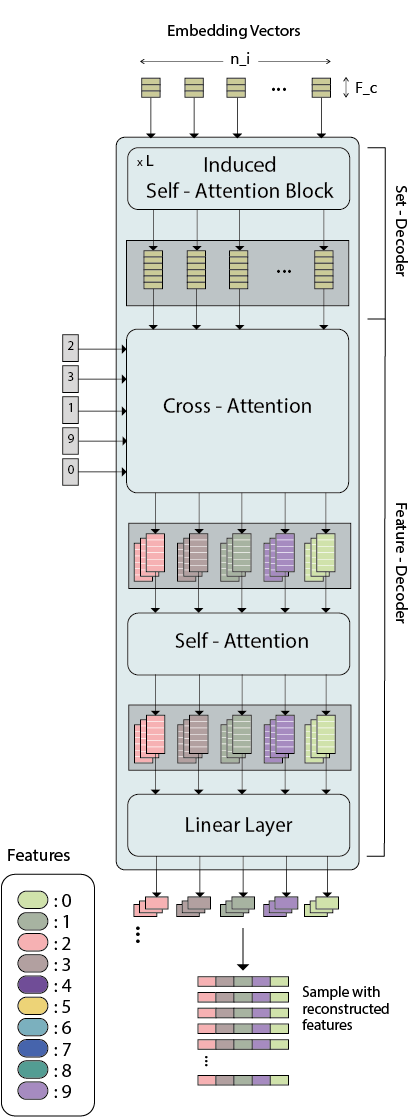}
   \caption{This figure illustrates the architecture of the decoder proposed for the \ac{MAE} experiments. Input are the embedding vectors, which are put into context to each other with ISAB layers. It follows cross attention with the feature encodings of the features that should be reconstructed. Finally, the learnt vectors each corresponding to one feature value are attended to each other and mapped to its final reconstructed scalar feature value.}
   \label{fig:mae-decoder}
\end{figure*}

Our Decoder proposed for the \ac{MAE} experiments can be divided into a set-decoder and a feature-decoder part as described in the main text. \cref{fig:mae-decoder} shows a detailed illustration of the two parts of the decoder.

\section{Induced Self-Attention Block}
\begin{figure*}
  \centering
   \includegraphics[width=0.45\linewidth]{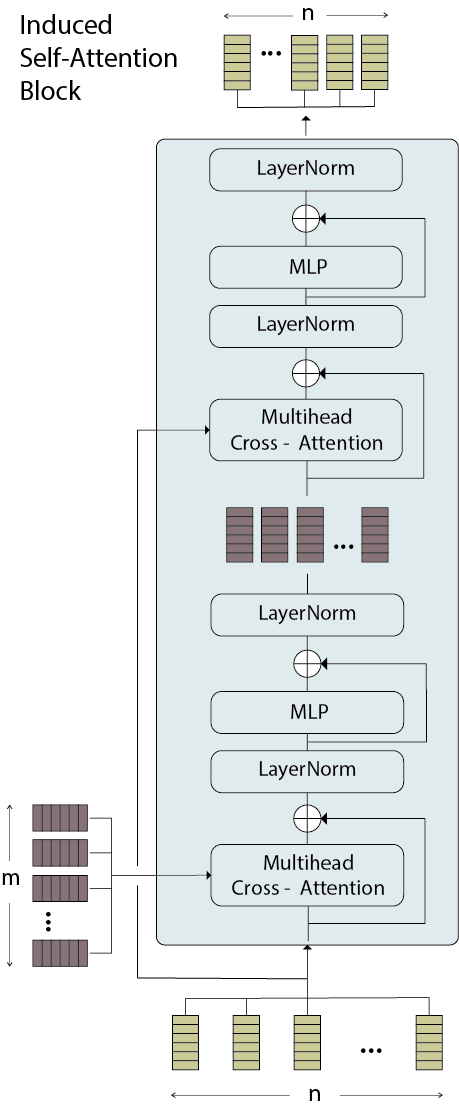}
   \caption{Illustration of the ISAB layer with an input set of $n$ elements and $m$ induced points \cite{lee2019set}. Plus sign indicates additive skip connections.}
   \label{fig:isab}
\end{figure*}

We use the ISAB layers as proposed in \cite{lee2019set}. \cref{fig:isab} shows an illustration of the detailed architecture, where $n$ denotes the number of input set elements and $m$ the number of learnt queries i.e. induced points.

\section{General Embedding of the FATE-MAE}
\begin{figure*}
  \centering
   \includegraphics[width=0.9\linewidth]{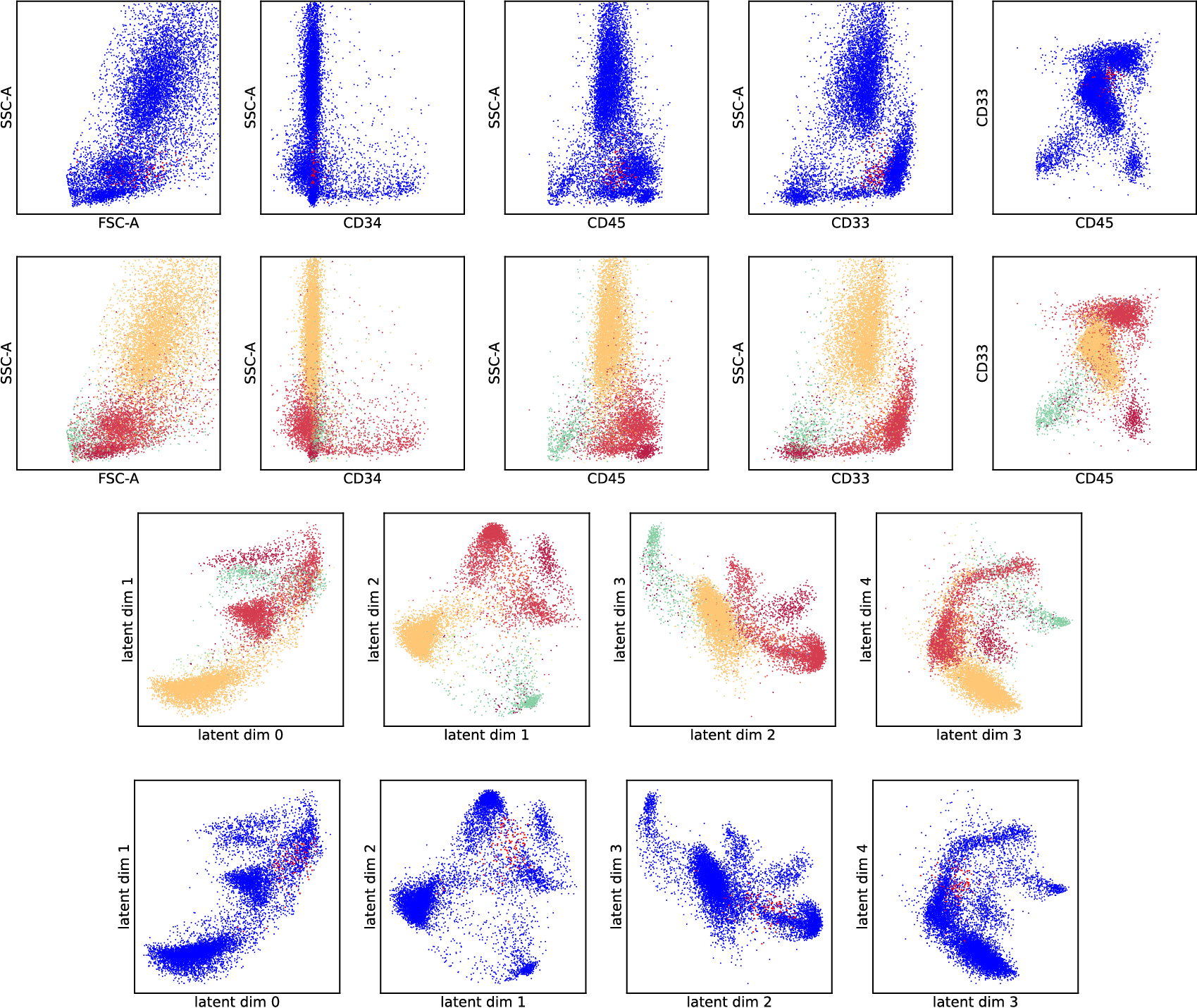}
   \caption{Visualization of a sample of Patient C from the AML-MRD dataset in its original feature space (row 1-2) and in the learnt general embedding space (row 3-4). Blue denotes healthy cells, red cancerous cells. The colours in row 2 and 3 indicate clusters of cell populations.}
   \label{fig:emgedding1}
\end{figure*}

\begin{figure*}
  \centering
   \includegraphics[width=0.9\linewidth]{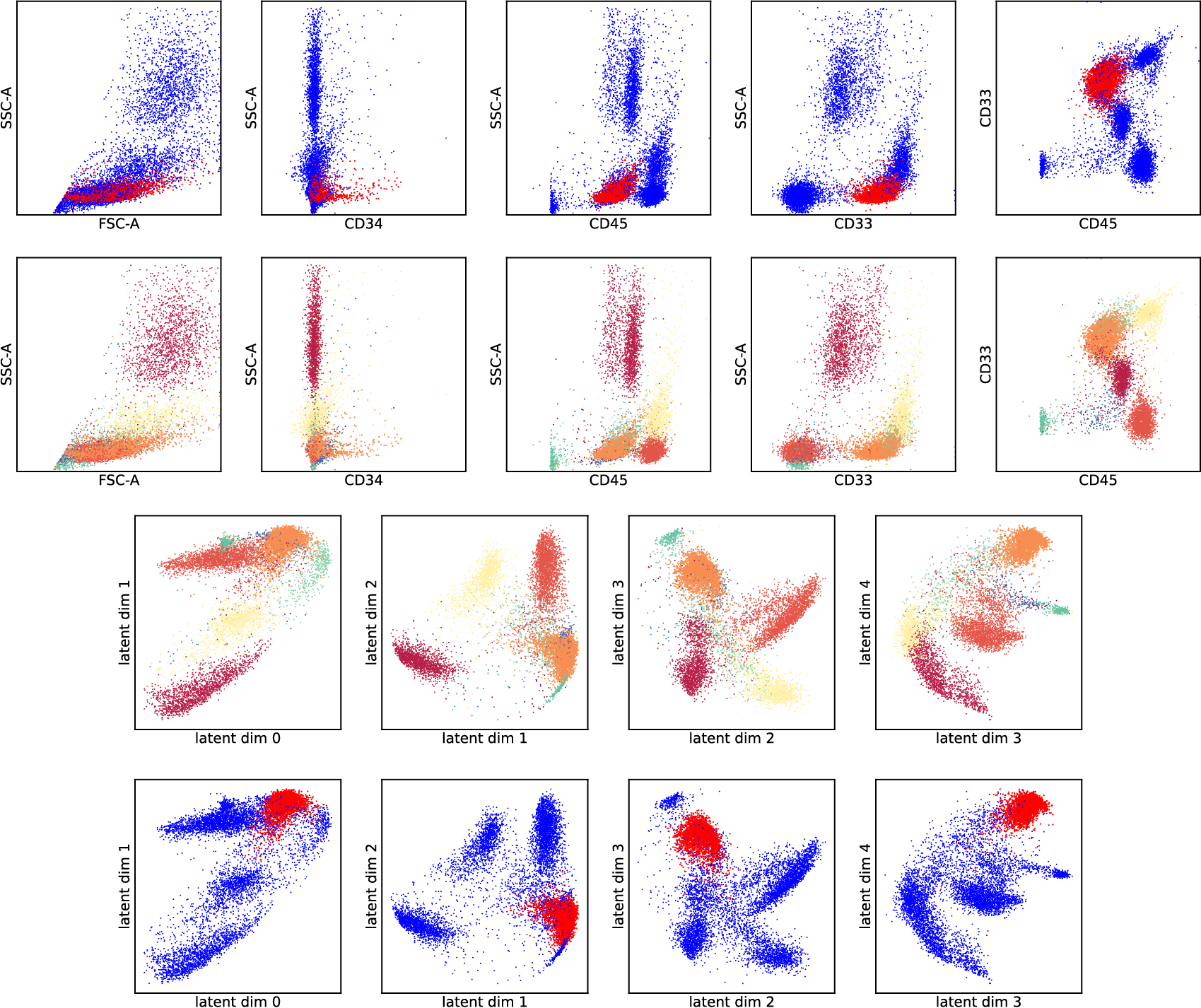}
   \caption{Visualization of a sample of Patient D from the AML-MRD dataset in its original feature space (row 1 -2) and in the learnt general embedding space (row 3-4). Blue denotes healthy cells, red cancerous cells. The colours in row 2 and 3 indicate clusters of cell populations.}
   \label{fig:embeddign2}
\end{figure*}

\cref{fig:emgedding1} and \cref{fig:embeddign2} visualize the embedding of two different samples generated with the pretrained FATE-MAE with the CONTROL, DIA and ALL-MRD dataset with a masking ratio of $0.25\%$. We visualize healthy (blue) versus cancerous (red) cells (row 1 and 4) as well as clustering of cell populations (row 2 and 3). The figures show that the learnt embedding forms meaningful clusters of cells with respect to clusters of the original feature space.


\end{document}